\documentclass[sigconf]{acmart}
\usepackage{xspace}
\usepackage{longtable}
\usepackage{array}
\usepackage{graphicx}
\usepackage{multirow}
\usepackage{tabularx}
\usepackage{enumitem}

\AtBeginDocument{%
  }

\copyrightyear{2026}
\acmYear{2026}
\setcopyright{cc}
\setcctype{by}
\acmConference[CHASE'26]{International Workshop on Cooperative and Human Aspects of Software }{April 12--18, 2026}{Rio de Janeiro, Brazil}
\acmBooktitle{International Workshop on Cooperative and Human Aspects of Software (CHASE'26), April 12--18, 2026, Rio de Janeiro, Brazil}
\acmPrice{}
\acmDOI{10.1145/3794860.3794912}
\acmISBN{979-8-4007-2484-8/2026/04}

\begin{document}

\title{Regression Testing in Remote and Hybrid Software Teams: An Exploratory Study of Processes, Tools, and Practices}


\author{Juliane Pascoal}
\email{jgp2@cesar.school}
\orcid{}
\affiliation{%
  \institution{CESAR School}
  \city{Recife}
  \country{Brazil}
}

\author{Cleyton Magalhaes}
\email{cleyton.vanut@ufrpe.br}
\orcid{}
\affiliation{%
  \institution{UFRPE}
  \city{Recife}
  \state{Pernambuco}
  \country{Brazil}}

\author{Ronnie de Souza Santos}
\email{ronnie.desouzasantos@ucalgary.ca}
\orcid{0000-0003-3235-6530}
\affiliation{%
  \institution{University of Calgary}
  \city{Calgary}
  \state{Alberta}
  \country{Canada}}

\begin{abstract}
Remote and hybrid work have transformed how software development teams organize, communicate, and assure quality. This study investigates how regression testing is performed and experienced under these distributed conditions. Using qualitative interviews with twenty software professionals from diverse organizations, we analyzed how regression testing processes, tools, and coordination practices adapt to remote and hybrid environments. The results show that while the core phases of regression testing remain stable, their execution increasingly depends on documentation, automation, and tool integration to support asynchronous collaboration. Communication and coordination challenges were mitigated through standardized reporting, shared repositories, and traceability mechanisms that replaced informal co-located interactions. These findings reveal regression testing as a socio-technical practice shaped by the interaction between human collaboration and digital infrastructure. Our study contributes to understanding how software quality assurance evolves under remote conditions and offers practical implications for teams and organizations adopting hybrid work models.

\end{abstract}



\maketitle

\newcommand{\searchStrings}{3\xspace}
\newcommand{\topResults}{100\xspace}
\newcommand{\totalResults}{400\xspace}
\newcommand{\initialResults}{100\xspace}
\section{Introduction}
\label{sec:introduction}

The increasing prevalence of remote and hybrid work in software development has reconfigured how teams plan, execute, and validate software systems. Over the past decade, distributed and hybrid development has expanded substantially, compelling organizations to reorganize workflows, strengthen communication mechanisms, and adapt quality assurance practices to sustain productivity and reliability. Recent studies show that geographically distributed collaboration has become a persistent feature of the software industry, with hybrid models now representing a dominant organizational form \cite{ralph2020pandemic,khanna2024hybrid,de2024post,smite2022future,christensen2025hybrid}. The COVID-19 pandemic further accelerated this transformation, forcing software organizations that previously relied on co-located development to adopt remote and hybrid modes of collaboration \cite{jansen2024remote,de2022grounded,nguyen2024work}.

Within this evolving context, software testing remains a central component of quality assurance. Among testing activities, regression testing is particularly critical, as it ensures that new modifications to the codebase do not reintroduce previously corrected defects or compromise existing functionality. As software development increasingly adopts agile methodologies and continuous integration pipelines, regression testing assumes an even more strategic role by supporting rapid feedback and iterative validation cycles \cite{tomar2022regression,rahmani2021empirical,wang2023test}. Sommerville \cite{sommerville2011software} defines regression testing as the incremental execution of test cases throughout the software lifecycle to verify that ongoing modifications do not introduce new faults. In agile and DevOps environments, this iterative process aligns closely with incremental development and facilitates the early detection of quality regressions \cite{samad2021regression,rosero2022software,greca2023state}.

Despite its importance, regression testing is inherently costly. It requires repeated execution of large test suites, often demanding substantial time, human effort, and computational resources. Studies consistently report that regression testing can consume between 50\% and 80\% of the total verification or maintenance budget, making it one of the most resource-intensive activities in software engineering \cite{rosero2022software,samad2021regression,greca2023state,tomar2022regression}. While these costs are considerable, regression testing remains indispensable for sustaining system stability, particularly in projects characterized by frequent releases and evolving requirements \cite{rahmani2021empirical,wang2023test}.

The transition to remote and hybrid work has reshaped how software testing activities are organized and executed \cite{jansen2024remote,de2024post,khanna2024software}. Teams that once relied on in-person coordination have had to adapt to asynchronous communication, distributed infrastructure, and cross-time-zone collaboration. Empirical studies highlight that these shifts introduce challenges related to communication latency, tool interoperability, infrastructure access, and coordination of testing tasks across geographically dispersed teams \cite{ralph2020pandemic,de2022grounded,christensen2025hybrid,nguyen2024work}. Limited in-person interaction can affect trust, cohesion, and feedback loops, factors that are fundamental for maintaining testing quality and timely defect resolution \cite{smite2022future,de2022grounded,khanna2024software}.

For regression testing in particular, these conditions raise questions about how processes, tools, and coordination mechanisms are adapted to remote contexts. Testing teams may experience disruptions in planning, execution, and reporting; automation pipelines may depend on infrastructure unevenly distributed across team members; and asynchronous communication may affect test traceability and defect management. At the same time, the adoption of cloud-based and collaborative testing tools has sought to mitigate some of these difficulties, though their effectiveness varies across organizations \cite{khanna2024software,christensen2025hybrid,de2024hybrid}. 

Existing literature suggests that while remote and hybrid models offer flexibility and potential gains in documentation and autonomy, they also introduce socio-technical frictions that alter traditional testing practices \cite{de2022grounded,smite2022future,de2024post,nguyen2024work}. Understanding how regression testing adapts within this reconfigured environment is therefore essential to inform both research and practice in software quality assurance. Such understanding can guide organizations in selecting tools, structuring communication protocols, and optimizing testing processes under distributed conditions.

To address these issues, this study adopts an exploratory perspective and investigates how regression testing practices evolve when executed by remote and hybrid teams. Rather than presupposing linear causality, the research aims to characterize and interpret variations in testing processes, tool use, and team coordination under distributed arrangements. 

Accordingly, the overarching research question (RQ) guiding this work is:

\textit{RQ. How does the adoption of remote and hybrid work arrangements shape the processes, tools, and socio-technical practices involved in regression testing?}

From this general question, three specific research questions are derived:

\textit{RQ1. How are regression testing processes currently organized and executed in remote and hybrid software development environments?}

\textit{RQ2. In what ways do testing tools influence or constrain the execution of regression testing in distributed contexts?}

\textit{RQ3. What contextual, organizational, and human factors characterize regression testing practices in remote and hybrid teams?}

This study employs a qualitative methodological design, drawing on interviews with professionals engaged in regression testing across distributed settings. The objective is to identify empirical patterns, challenges, and adaptations that characterize the execution of regression testing under remote and hybrid conditions, contributing to a more comprehensive understanding of software quality assurance in contemporary development environments. The remainder of this paper is organized as follows. Section~\ref{sec:background} reviews prior research on regression testing and distributed software development. Section~\ref{sec:method} describes the methodological approach, including data collection and analysis. Section~\ref{sec:results} presents the empirical findings. Section~\ref{sec:discussion} discusses the implications for software engineering research and practice. Section~\ref{sec:conclusions} concludes the paper and identifies directions for future work.

\section{Background}
\label{sec:background}
This section introduces the core concepts of the study, addressing regression testing as a key mechanism for maintaining software quality, examining how remote and hybrid work have reshaped software engineering practices, and identifying a research gap concerning how regression testing adapts within distributed development contexts.

\subsection{Regression Testing}

Regression testing is a central component of software quality assurance and serves to verify that new modifications to a system do not compromise previously validated functionality. It involves the repeated execution of test suites to ensure that code changes do not introduce unintended effects or degrade existing behavior \cite{tomar2022regression,rahmani2021empirical,wang2023test}. As discussed by Sommerville \cite{sommerville2011software}, regression testing comprises a set of test cases that are developed and executed incrementally as the program evolves. This practice aligns with agile and iterative development, where frequent releases require constant validation of core system behavior \cite{rosero2022software,samad2021regression,greca2023state}.

The literature identifies regression testing as a systematic mechanism for maintaining software reliability throughout its lifecycle. Empirical studies describe it as a process that supports ongoing system evolution by detecting whether new functionality interferes with previously stable components \cite{tomar2022regression,rahmani2021empirical,greca2023state}. Reviews of regression testing research show that it has evolved from an ad-hoc validation practice into a structured activity characterized by distinct methods for test-case selection, prioritization, and minimization \cite{rosero2022software,samad2021regression,wang2023test}. These strategies collectively aim to improve fault detection effectiveness and optimize the cost of repeated testing cycles.

Despite its indispensability, regression testing is widely recognized as one of the most resource-intensive phases of software development. It demands considerable computational and human effort, as large test suites must be executed multiple times to ensure the stability of modified systems \cite{rosero2022software,samad2021regression,greca2023state,tomar2022regression}. Studies of industrial practice consistently report that regression testing can consume a significant portion of the overall testing or maintenance budget, confirming its substantial operational cost \cite{rosero2022software,samad2021regression,greca2023state}. In response to these constraints, recent work has proposed automated regression testing techniques that reduce execution time without compromising coverage \cite{rahmani2021empirical,wang2023test}. Nevertheless, researchers agree that regression testing remains a foundational activity for sustaining reliability and preventing fault reintroduction in projects that follow agile or continuous-integration models \cite{rosero2022software,samad2021regression,greca2023state,wang2023test}.

\subsection{Remote and Hybrid Work}
The organization of software development teams has undergone profound transformation with the rise of remote and hybrid work. Research on distributed and global software development has long examined geographically and temporally dispersed collaboration, but recent years have witnessed a consolidation of hybrid models as a dominant mode of operation \cite{ralph2020pandemic,khanna2024hybrid,smite2022future}. The COVID-19 pandemic accelerated this shift, forcing even traditionally co-located organizations to restructure their workflows and adopt remote collaboration practices \cite{jansen2024remote,de2022grounded,nguyen2024work}. These transformations required teams to adapt communication channels, coordination mechanisms, and quality assurance processes to maintain productivity under distributed conditions \cite{de2024post,christensen2025hybrid,de2024hybrid}.

Studies describe remote and hybrid work as phenomena that combine flexibility and global reach with new organizational and socio-technical challenges. While distributed arrangements enable autonomy and access to diverse expertise, they also complicate collaboration and knowledge sharing due to reduced visibility and limited informal communication \cite{ralph2020pandemic,de2022grounded,nguyen2024work}. Empirical research has shown that geographical dispersion can hinder the establishment of trust and shared understanding among team members and that asynchronous communication may delay feedback and defect resolution \cite{smite2022future,christensen2025hybrid,alharbi2022empirical}. In the context of software testing, where verification and validation rely on synchronized execution and timely feedback, these factors can significantly affect quality and coordination \cite{jansen2024remote,de2024post}.

The software industry continues to explore how to institutionalize sustainable practices for remote and hybrid collaboration. Many processes that were previously standardized for co-located development now require continuous experimentation and adaptation \cite{khanna2024software,de2024post,smite2022future}. Empirical accounts from industry and academia emphasize that effective hybrid work depends on robust communication infrastructure, consistent tool usage, and deliberate coordination strategies \cite{jackson2022collaboration,de2024hybrid,christensen2025hybrid}. Insufficient alignment between tools, practices, and organizational expectations can impair testing reliability and overall software quality \cite{ralph2020pandemic,de2022grounded,alharbi2022empirical}. As hybrid work becomes a defining feature of software engineering, organizations continue to balance autonomy, accountability, and cohesion in reconfigured socio-technical environments \cite{nguyen2024work,de2024post,khanna2024software}.

\subsection{Remote and Hybrid Work in the Context of Software Testing}

Remote and hybrid work have transformed how testing activities are coordinated and executed in software development projects. Empirical research conducted after the pandemic describes how testing professionals needed to adjust communication routines, tool usage, and access to infrastructure to maintain effective quality assurance \cite{jansen2024remote,alharbi2022empirical}. Studies focusing on developers and testers working remotely indicate that defect tracking, feedback exchange, and regression cycles became increasingly dependent on asynchronous communication and digital collaboration platforms. This shift has altered the rhythm of testing processes, reinforcing the importance of documentation, visibility, and traceability to sustain quality in distributed environments.

Evidence from investigations on distributed software testing shows that remote work has introduced challenges related to synchronization, infrastructure stability, and tool integration \cite{alharbi2022empirical,jansen2024remote}. Researchers have observed that testers experienced delays in coordination, difficulties accessing shared environments, and additional effort to align testing schedules across time zones. These studies also suggest that communication among distributed testers and developers became more formalized, with written channels replacing informal in-person exchanges \cite{jansen2024remote}. Although this change supported accountability and traceability, it also reduced immediacy in problem solving and slowed defect resolution cycles that were previously facilitated by co-located collaboration.

Hybrid work has also influenced how organizations manage testing infrastructure and automation. Industrial studies describe a gradual movement toward continuous integration, automated pipelines, and cloud-based testing environments as teams adapt to working from multiple locations \cite{de2022grounded, de2024post}. These infrastructures allow testing tasks to proceed asynchronously and help ensure consistency in quality assurance despite physical separation. However, their effective use depends on coordinated configuration, tool interoperability, and shared process standards across geographically distributed teams. Hybrid arrangements therefore require deliberate organizational strategies to maintain coherence between automation systems, human workflows, and communication practices.

Mapping and global studies on hybrid software development indicate that testing and quality assurance have been among the areas most affected by these structural changes \cite{khanna2024hybrid,nguyen2024work}. Testing professionals report that remote and hybrid conditions have extended testing cycles, increased reliance on automation, and heightened the need for collaborative discipline. While hybrid work enhances flexibility and autonomy, it also demands reliable infrastructure, standardized procedures, and continuous communication to preserve alignment across roles and locations.

\subsection{Research Opportunity}

Existing research has explored regression testing and remote or hybrid work as separate topics, but their intersection remains under-investigated. The literature provides only partial understanding of how regression testing processes, tools, and communication practices evolve when distributed across locations. This limitation is increasingly relevant because activities that were once performed through co-located interaction, including test planning, debugging, and validation, are now executed through asynchronous collaboration and digital platforms. These changes have introduced new dependencies on automation pipelines, shared repositories, and structured communication routines that compensate for the absence of physical proximity. Investigating regression testing within remote and hybrid contexts represents a timely opportunity to advance knowledge about how software teams adapt socio-technical practices to sustain reliability, traceability, and efficiency in geographically dispersed environments. Such inquiry can also inform organizations seeking to strengthen quality assurance processes under contemporary models of distributed software development.

\section{Method}
\label{sec:method}

This study investigates how regression testing is performed and
experienced in remote and hybrid software development teams.
The objective is to understand whether remote work has influenced
regression testing processes, tools, and collaboration, and to iden-
tify the adaptations practitioners have made. We used a qualitative
interview design \cite{seaman1999qualitative,hove2005experiences}
organized into four sequential phases: (1)
qualitative data collection through structured interviews \cite{hove2005experiences}, (2) thematic analysis \cite{cruzes2011recommended}, (3) synthesis of findings, and (4) validation through member checking \cite{santos2017member}. The methodological framework followed the
structure proposed by \cite{ralph2020empirical} and ensured rigor through systematic
data collection, analysis, and verification.

\subsection{Participant Definition and Sampling}

Participants were software professionals with direct experience in executing regression testing under both remote and on-site arrangements. Selection criteria considered (a) seniority level (junior, mid-level, senior), (b) work modality (remote and on-site), (c) client geography (regional, national, international), (d) testing platforms (web, mobile, desktop, virtual and augmented reality), and (e) testing approach (manual and automated). These dimensions were used to define a participant profile aligned with the investigative nature of the study and are consistent with participant diversity practices recommended for qualitative empirical software engineering research \cite{seaman1999qualitative,lenberg2024qualitative,ralph2020empirical}.

Initial recruitment identified twelve participants meeting these criteria. A snowball sampling strategy~\cite{seaman1999qualitative} was then employed: each interviewee referred other qualified professionals, and the process continued in blocks of four interviews until theoretical saturation was achieved. The final sample comprised twenty professionals with three to fifteen years of experience in software testing, representing diverse organizations and domains.

\subsection{Instrument Design and Piloting}

Data were collected through structured interviews supported by a standardized guide \cite{hove2005experiences,seaman1999qualitative}. The instrument was organized into three thematic blocks derived from the research questions:

\begin{enumerate}[label=(\alph*)]
\item \textbf{Process:} How regression testing is planned and executed in organizations.
\item \textbf{Tools:} How tools influence and support regression testing.
\item \textbf{Context:} How remote or hybrid work conditions affect regression testing practices.
\end{enumerate}

\begin{table}[ht]
  \caption{Interview Guide}
  \label{tab:interview}
  \footnotesize
  \vspace{-8px}
  \begin{tabularx}{\linewidth}{p{1.7cm} X}
    \toprule
    Section & Questions \\
    \midrule

    Process &
    1. What criteria are considered when deciding whether to perform regression testing manually or automatically? \\
    & 2. Could you outline the sequence of steps followed in regression testing, from planning to completion? \\
    & 3. Which steps or criteria in the regression process could be optimized? \\
    & 4. At what points in the project is regression testing executed? \\
    & 5. What criteria guide the selection of test cases used in regression testing? \\
    & 6. What stability or entry conditions are essential before initiating regression testing? \\
    & 7. How is the team divided when regression testing is required? \\
    & 8. What is the average time reserved for regression testing? \\
    \midrule

    Tools &
    9. What tools are used for regression testing? \\
    & 10. What criteria guide the decision to use or not use tools in regression testing? \\
    & 11. In which testing approaches are these tools most applied? \\
    & 12. What are the main advantages of the tools used? \\
    & 13. What are their disadvantages or limitations? \\
    & 14. Do you have recommendations to mitigate tool-related issues? \\
    & 15. Have tools ever hindered regression test execution? \\
    \midrule

    Remote and Hybrid Context &
    16. How has remote work affected the execution of regression tests in your project? \\
    & 17. What benefits have you observed when performing regression testing remotely? \\
    & 18. What difficulties have you encountered under remote work conditions? \\
    & 19. Have communication failures ever disrupted regression testing? \\
    & 20. Have cooperation problems within the team affected regression testing? \\
    & 21. Compared to on-site work, has remote work changed the time required for regression testing? \\
    & 22. In remote contexts, have escaped bugs increased or decreased? \\
    & 23. Have test-case selection criteria ever hindered regression testing? \\
    & 24. Have escaped bugs been detected after regression that could have been avoided? \\
    & 25. How can regression testing be useful for onboarding new QA team members? \\
    
    \bottomrule
  \end{tabularx}
  \vspace{-10px}
\end{table}

Each block included open-ended questions with optional probes to elicit detailed accounts of practices and experiences. The guide was validated through a \textit{pre-test} with two participants possessing prior experience in regression testing and remote teamwork. These pilot interviews (approximately 50 minutes each) ensured that the questions were clear, non-redundant, and elicited reflective responses \cite{hove2005experiences}. Minor revisions were made to the wording and order of questions to improve flow. The refined interview guide was then finalized for the main study.

Prior to the interviews, potential participants received an email titled \textit{“Participant and Project Categorization”}, which included screening questions about education, experience, certifications, gender (open text for inclusivity), testing role, client location, and exposure to remote projects. Responses were used to confirm eligibility and ensure diversity. Approved participants then proceeded to the interview stage.

\subsection{Data Collection Procedures}

Interviews were conducted individually using Google Meet between January 2024 and April 2025. Each session lasted between thirty and fifty minutes and followed the finalized script. Participants granted written consent for recording. The researcher began each session by explaining the study's purpose, confidentiality measures, and the voluntary nature of participation \cite{ralph2020empirical,lenberg2024qualitative}.  
All interviews were recorded as video, converted to audio, and then transcribed for analysis. Each transcript was manually reviewed against the audio for accuracy. The team maintained supplementary notes during and after each interview to capture contextual observations and reflections. Interviews were concluded when theoretical saturation was reached, meaning that no new insights were emerging and discussions began to repeat prior themes. Signs of saturation were first observed around interview fourteen, confirmed by interview seventeen, yet data collection continued until twenty interviews had been completed to ensure comprehensive coverage \cite{ralph2020empirical}. This process resulted in a verified corpus of twenty transcripts ready for analysis.

\subsection{Data Analysis}

Data were analyzed using thematic analysis~\cite{cruzes2011recommended,seaman1999qualitative}, following established guidance for qualitative research in software engineering \cite{ralph2020empirical,lenberg2024qualitative}. The process involved four main stages:

\begin{enumerate}[label=(\alph*)]
\item \textbf{Data Preparation:} Each research question and its sub-questions were mapped to participants’ coded responses to structure the dataset.
\item \textbf{Coding:} Key text segments from transcripts were highlighted and labeled with descriptive codes representing distinct ideas or experiences. Each code was color-marked to maintain traceability across participants.
\item \textbf{Categorization:} Related codes were grouped into categories representing broader conceptual similarities. Coding and categorization were performed manually using Microsoft Excel, as this provided adequate control and transparency for the dataset size.
\item \textbf{Synthesis:} Relationships between categories were analyzed to identify overarching patterns and to develop themes that explained how regression testing practices evolved under remote and hybrid conditions.
\end{enumerate}

Manual tabulation enabled cross-referencing of participant statements and efficient visualization of thematic relationships. The iterative movement between codes and categories ensured conceptual coherence and data-driven interpretation, consistent with prior recommendations for rigorous thematic synthesis in software engineering research~\cite{cruzes2011recommended,lenberg2024qualitative,ralph2020empirical}. Coding was independently reviewed by two researchers and subsequently discussed until agreement was achieved, ensuring consistency and shared understanding of the emerging categories. The analysis continued until code repetition across interviews confirmed saturation.

\subsection{Member Checking and Validation}

Following the analysis, a validation phase was conducted to verify the credibility of interpretations. A structured questionnaire was sent to all twenty participants, summarizing the preliminary findings across the three axes—process, tools, and remote context. Participants rated their level of agreement on a five-point Likert scale (from “strongly agree” to “strongly disagree”) and could provide written comments. Seventeen responses were received.

Feedback confirmed that participants recognized their experiences in the synthesized results. Minor adjustments were made to improve precision and wording. This step functioned as member checking \cite{santos2017member}, ensuring that the derived interpretations were faithful to participants’ intended meanings and aligned with standard credibility verification practices in qualitative software engineering research \cite{lenberg2024qualitative,ralph2020empirical}.

\subsection{Ethical Considerations}

All procedures complied with ethical research principles. Participants were informed about study objectives, data handling, and confidentiality. Identifiable information was excluded from transcripts, and pseudonyms were used in all analyses and reports. Only the research team had access to recordings and transcripts, which are securely stored on institutional servers. Participation was voluntary, and no incentives were provided. Ethical conduct in software engineering research emphasizes participant consent, privacy protection, and transparency throughout the study process \cite{ralph2020empirical,lenberg2024qualitative,seaman1999qualitative}.

\section{Results}
\label{sec:results}

This section presents the main findings of the study according to the three research questions. The analysis reveals how regression testing practices are organized and executed in remote and hybrid environments, how tools shape these processes, and how socio-technical factors affect testing coordination and outcomes.

\subsection{Participant Demographics} \label{sec:demographics} The study included twenty software quality assurance professionals involved in regression testing across different organizations. As shown in Table~\ref{tab:Demographics}, participants represented varied levels of experience and seniority. Most respondents were mid-level professionals (\textit{pleno}, $n=10$), followed by senior ($n=4$), specialist ($n=3$), and junior testers ($n=3$). In terms of professional experience, eight participants had between four and seven years of practice, six had between seven and ten years, four had more than ten years, and two had between one and three years. The sample was predominantly male ($n=13$), with seven female participants. Regarding client regions, seven participants reported working primarily with clients in North America, six in South America, six in Europe, and one in Asia. Overall, the sample reflects a population of mid-career QA professionals, mostly male, with intermediate to advanced experience and distributed collaboration with clients across multiple continents.

\begin{table}[ht]
\centering
\caption{Participant Demographics (N=20)}
\renewcommand{\arraystretch}{1}
\label{tab:Demographics}
\begin{tabular}{llr}
\hline\noalign{\smallskip}

\multirow{2}{*}{ \textbf{Gender} } 
& Male & 13 individuals \\
& Female & 7 individuals \\ \midrule

\multirow{4}{*}{ \textbf{Seniority Level} } 
& Junior & 3 individuals \\
& Mid-level & 10 individuals \\
& Senior & 4 individuals \\
& Specialist & 3 individuals \\ \midrule

\multirow{4}{*}{ \textbf{Experience (Years)} } 
& 1–3 years & 2 individuals \\
& 4–7 years & 8 individuals \\
& 7–10 years & 6 individuals \\
& More than 10 years & 4 individuals \\ \midrule

\multirow{4}{*}{ \textbf{Client Region} } 
& North America & 7 individuals \\
& South America & 6 individuals \\
& Europe & 6 individuals \\
& Asia & 1 individual \\

\noalign{\smallskip}\hline
\end{tabular}
\end{table}

\subsection{RQ1. Organization and Execution of Regression Testing in Remote and Hybrid Environments}

Interviews indicate that regression testing in remote and hybrid teams generally follows a structured process composed of three main phases: \textit{planning}, \textit{execution}, and \textit{closure}. These phases remain similar to traditional in-person workflows but have been adapted to support asynchronous coordination, virtual communication, and distributed tool usage.

\textbf{Planning.}  
Participants described the planning phase as the foundation for organizing regression testing in a distributed setting. It focuses on ensuring that all prerequisites are completed and that information is shared before execution begins. Activities include confirming that development stories are finished, analyzing documentation, identifying code changes and affected services, creating and reviewing test cases, and preparing the test environment.  
P2 emphasized that readiness depends on the prior completion of development tasks: \textit{“To start the regression cycle, we make sure the story is ready; only then can we plan the tests.”}  
For many participants, documentation analysis replaced the informal exchanges that occurred in co-located work. P3 explained: \textit{“First, we analyze the documentation to verify if the criteria are within the expected parameters.”}  
Others highlighted the importance of scope definition and communication in remote contexts. P13 noted: \textit{“We analyze what will be delivered in this version and which functionalities that we already delivered will be impacted.”} P1 added: \textit{“I check which services were affected by that change to make sure other systems will not be impacted.”}  
Environmental preparation also took on greater importance in remote and hybrid teams where physical access to shared infrastructure is limited. P3 described: \textit{“We prepare the devices, set up the language, the server, the binaries, and everything that will be used.”}  
These accounts show that in remote and hybrid arrangements, the planning phase depends on structured documentation and traceability rather than spontaneous coordination, reflecting the need for clarity before asynchronous execution begins.

\textbf{Execution.}  
During execution, participants reported performing both automated and manual tests, depending on available infrastructure and project priorities. Many highlighted that the lack of physical proximity required more deliberate coordination and tool-based monitoring.  
Automation was frequently used for repetitive validations, as P6 described: \textit{“Before deploying to production, we have a regression sprint where we execute the entire suite; what is automated, we run and record the result.”}  
However, manual testing remained essential for functions that required direct human verification or when infrastructure limitations affected automation. P20 explained: \textit{“In the execution phase, we work manually using Azure DevOps; we create a test cycle with the mapped cases and execute them remotely.”}  
Participants also mentioned new coordination challenges, particularly related to pipelines and asynchronous troubleshooting. P1 shared: \textit{“Sometimes someone pushes a change, and a functional test that worked before breaks; we identify the failure and fix the pipeline remotely.”}  
Managing and classifying errors was part of this same distributed cycle. P5 stated: \textit{“We collect the errors and check if it was a script failure or an application failure.”}  
The need to retest defects after distributed corrections was also emphasized. P20 remarked: \textit{“We only move to production when the bug is retested; we run a new cycle only with the tests that failed in the previous regression.”}  
These excerpts reveal that the execution phase relies heavily on continuous integration tools, asynchronous communication, and self-managed coordination, replacing the in-person observation typical of co-located testing.

\textbf{Closure.}  
The closure phase formalizes results and ensures shared visibility across geographically distributed stakeholders. Participants reported recording outcomes in collaborative tools such as JIRA, Confluence, and Azure DevOps, producing reports, and generating release versions.  
P9 described: \textit{“At the end, I enter the results in JIRA; that is the final step of my tests.”}  
P5 highlighted that this reporting replaces the informal verbal updates of co-located work: \textit{“After we conclude the cycle, we collect the information, prepare the report with all evidence, and send it to the client.”}  
P13 reinforced the role of integrated platforms for transparency: \textit{“Within Azure we have a report that records all test cycles and failures, so everything is managed inside the same tool.”}  
Through these practices, distributed teams maintain visibility and accountability even without shared physical spaces.

Overall, regression testing in remote and hybrid environments preserves the traditional methodological structure of planning, execution, and closure, but it is more dependent on digital tools, explicit documentation, and disciplined coordination. As P13 summarized: \textit{“We start the testing cycle by preparing the system, setting up devices and servers to ensure everything is ready, since in remote work we depend on these setups being accessible to everyone.”} Teams compensate for reduced in-person interaction by strengthening documentation, standardizing procedures, and relying on collaborative platforms to ensure alignment and continuity across locations.

\subsection{RQ2. Influence of Tools on Regression Testing Processes}

The study identified a diverse ecosystem of tools that support regression testing. These tools are organized into five categories: \textit{test management}, \textit{automation}, \textit{execution}, \textit{development}, and \textit{continuous integration}. Tool selection is shaped by project context, team maturity, and organizational infrastructure.

\textbf{Test management.} JIRA was the most widely used tool, often combined with Confluence, Zephyr, or Excel spreadsheets. Participants valued its traceability, integration with defect tracking, and visibility across teams but also mentioned its slowness and complex configuration. Azure DevOps was appreciated for centralizing documentation, test execution, and client visibility, although users noted limitations in real-time updates and interface usability. Manual tracking through spreadsheets remained common in smaller teams due to simplicity, despite concerns about loss of version control and traceability.

\textbf{Automation.} Commonly used frameworks include Cypress, PyTest, Robot, Behave, Playwright, and Appium. Participants highlighted the ease of scripting and learning in Cypress and the stability and community support of PyTest. However, automation required frequent configuration adjustments and sometimes faced compatibility and maintenance issues. Integration between automation frameworks and test management tools was often partial, reducing overall visibility of automated test results.

\textbf{Execution and development.} Tools such as Postman, RabbitMQ, and Insomnia supported API integration tests, with Postman being the most prevalent due to its fast execution and reporting capabilities. For development support, environments such as Visual Studio Code, PyCharm, and DBeaver facilitated code consistency and collaboration, while DataDog offered metrics for performance and defect tracking but incurred high costs.

\textbf{Continuous integration.} Jenkins and CircleCI were used to enable automated regression pipelines. Jenkins was recognized for its reliability and ability to support frequent releases. Participants did not report major disadvantages for these tools.

Across all categories, participants emphasized that the effectiveness of tools depends more on their alignment with team context, available infrastructure, and user proficiency than on the tools themselves. While tools provided visibility and standardization, they also introduced challenges related to configuration overhead, integration gaps, and learning curves that could hinder productivity in distributed settings.

\subsection{RQ3. Socio-Technical and Organizational Factors Shaping Regression Testing}

The transition to remote and hybrid work introduced significant changes in communication, collaboration, and infrastructure. These changes affected not only team performance but also the quality and reliability of testing outcomes.

\textbf{Communication and coordination.}  
Most participants identified communication as the most critical challenge under remote conditions. Several described difficulties arising from time-zone differences, asynchronous feedback, and inconsistent information between roles. As P1 explained, \textit{“I work with a global team in India, which is eight and a half hours ahead, and that delay affects our work a lot.”} P3 highlighted that miscommunication often required the team to redo tasks: \textit{“There were many communication failures; some people thought they were answering correctly, but the information was wrong, and the whole team had to redo the tests to deliver on time.”}  
Other participants mentioned that misunderstandings about test versions or assigned tasks invalidated entire test cycles. P9 recalled, \textit{“I had already started the regression tests, and someone merged new code into the version. When I realized the version had changed, I was already finishing the tests.”} These accounts show that asynchronous communication and overlapping responsibilities often led to rework and coordination overhead. At the same time, the increased use of documentation and shared tools such as JIRA and Azure DevOps was viewed as a way to preserve alignment and traceability in distributed teams.

\textbf{Productivity and focus.}  
Despite coordination difficulties, many respondents reported greater concentration and efficiency while working remotely. The reduction of interruptions and the flexibility of home environments allowed testers to sustain longer, more focused work sessions. P7 commented, \textit{“I think performance even improved because people are more concentrated at home, in a quieter environment.”} Similarly, P4 noted the contrast with on-site work: \textit{“In the office, I had to stop all the time to help someone, and my tests would get delayed. At home, I was less distracted and performed better.”}  
Some participants associated this increased focus with a stronger sense of autonomy and comfort. As P17 described, \textit{“Working remotely makes me feel more comfortable; I think I can focus more at home.”} Overall, many interviewees perceived remote arrangements as enabling deeper engagement and higher personal productivity during regression cycles.

\textbf{Quality outcomes and learning.}  
Participants offered mixed perspectives on how remote work influenced defect rates and learning processes. Some observed more escaped bugs due to weakened communication and lack of peer verification. P19 mentioned, \textit{“When we were together, I would keep an eye on my colleague’s work, especially the junior tester, and that helped catch mistakes. Remotely, that interaction decreased.”}  
Others, however, reported improvements in quality due to greater focus and fewer distractions. P13 said, \textit{“I believe escaped bugs decreased. Remotely, we can monitor things more closely, identify problems faster, and solve them before they reach the client.”}  
Regression testing also emerged as a key learning activity for new QA members. P2 explained, \textit{“It helps new testers understand what already exists in the system; they go through flows they don’t know and gain knowledge about other functionalities.”} P10 added, \textit{“If a new person joins, they can already get a sense of how things work just by running the regression tests.”} These testimonies show that regression testing in remote and hybrid contexts supports both quality control and onboarding.

\textbf{Infrastructure and resource constraints.}  
Technical limitations were another recurring theme, including unstable VPN connections, limited access to test devices, and dependency on client data. P5 summarized, \textit{“Sometimes the VPN is slow, that’s all.”} P9 pointed out the lack of proper hardware: \textit{“I don’t have a physical device to test, only simulated ones, and that slows everything down.”}  
Similarly, P13 explained how missing equipment extended execution time: \textit{“I once lacked the necessary devices to validate a feature, and that delayed the process until I could find the right equipment.”}  
These resource constraints required creative adjustments, such as remote environment setups and asynchronous test handoffs between team members in different regions.

In summary, the findings reveal a dual and nuanced reality shaped by the adoption of remote and hybrid work. On one side, these arrangements amplify socio-technical challenges, particularly in communication and coordination, where asynchronous exchanges, time-zone differences, and uneven tool integration often lead to misunderstandings, rework, and dependency on stable infrastructure. On the other side, participants consistently emphasized the advantages of hybrid work, reporting higher productivity, greater concentration, and a stronger sense of autonomy when testing from home or alternating between environments. The data also indicate that regression testing in these distributed settings acquires a more structured and reflective character: it becomes document-driven, tool-mediated, and explicitly collaborative through shared repositories and traceable workflows. Overall, remote and hybrid work transform regression testing into a practice that, while facing communication and infrastructure constraints, benefits from improved focus, flexibility, and procedural rigor, sustaining software quality through balanced adaptation between human coordination and technological support. Table \ref{tab:rq3_summary} summarizes this evidence.

\begin{table}[ht]
  \caption{Summary of Socio-Technical and Organizational Findings}
  \label{tab:rq3_summary}
  \vspace{-8px}
  \begin{tabularx}{\linewidth}{p{2.5cm} X}
    \toprule
    Theme & Summary of Findings \\
    \midrule

    Communication and Coordination &
    Mostly negative. Asynchronous exchanges and time-zone differences often caused misunderstandings and rework, but documentation practices and tool-based traceability improved under hybrid settings. \\
    \midrule

    Productivity and Focus &
    Mostly positive. Remote and hybrid arrangements increased concentration, autonomy, and comfort, reducing interruptions and enabling faster regression cycles. \\
    \midrule

    Quality and Learning &
    Mixed (slightly positive). Some participants noted more escaped bugs due to reduced collaboration, while others saw fewer defects and valued regression testing as a learning mechanism for onboarding new QA members. \\
    \midrule

    Infrastructure and Resources &
    Mostly negative. VPN latency, unstable internet, and limited access to testing devices constrained execution, though teams adapted with remote setups and flexible coordination. \\

    \bottomrule
  \end{tabularx}
  \vspace{-10px}
\end{table}

\subsection{Answering the General Research Question}
\label{sec:rq_general}

The central research question guiding this study was: 
\textit{How does the adoption of remote and hybrid work arrangements shape the processes, tools, and socio-technical practices involved in regression testing?} 

The analysis across the three specific research questions reveals that remote and hybrid work have reshaped regression testing at multiple, interdependent levels, processual, technological, and social, introducing both challenges and adaptations that redefine how testing is organized and executed.

\textbf{Processes.}  
Findings from RQ1 demonstrate that regression testing continues to follow the traditional phases of \textit{planning}, \textit{execution}, and \textit{closure}, but each phase has evolved to accommodate distributed collaboration. Planning activities, once supported by in-person discussions, now rely heavily on detailed documentation and well-defined scope analysis to ensure shared understanding across time zones. Execution has become more asynchronous, requiring deliberate coordination and stricter alignment between automation and manual validation. Closure activities, such as reporting and evidence generation, are now centralized in digital platforms to provide visibility and accountability to geographically dispersed teams. Overall, regression testing has become more structured and document-driven, compensating for the loss of spontaneous communication through procedural rigor and traceable workflows.

\textbf{Tools.}  
As shown in RQ2, tools play a central role in enabling regression testing under remote and hybrid conditions. Test management platforms such as JIRA, Zephyr, and Azure DevOps act as coordination hubs that integrate planning, execution, and reporting. Automation frameworks like Cypress, PyTest, and Robot facilitate continuous validation, while Jenkins and CircleCI sustain integration and deployment pipelines across distributed infrastructures. However, participants also noted challenges such as steep learning curves, configuration overhead, and partial integration among tools. These limitations reveal that tool adoption is not purely technical but contingent on context, team maturity, and user expertise. In hybrid teams, effective tool use depends on balancing automation with manual oversight, ensuring that both human judgment and technological mediation are harmonized to sustain quality assurance.

\textbf{Socio-technical practices.}  
Results from RQ3 highlight that the transition to remote and hybrid work has transformed how teams coordinate, communicate, and share responsibility during regression testing. Communication emerged as both a challenge and an adaptation: while asynchronous channels and time-zone differences created delays and misunderstandings, teams compensated by documenting more systematically and relying on shared repositories for traceability. Productivity and focus were generally perceived to improve under remote and hybrid arrangements, as testers experienced fewer interruptions and greater autonomy. At the same time, infrastructure limitations, such as VPN latency and restricted access to devices, remained recurrent barriers. Importantly, regression testing also acquired a pedagogical role in distributed contexts, serving as a mechanism for onboarding and skill development among new QA members. These socio-technical adaptations suggest that regression testing has become not only a quality assurance activity but also a collaborative and learning-oriented practice.

\textbf{Synthesis.}  
Overall, the evidence indicates that remote and hybrid work have not merely displaced regression testing processes to virtual environments but have redefined them through increased reliance on tools, documentation, and distributed coordination. The process has evolved toward greater formalization and automation, while still depending on human expertise to interpret results and maintain cohesion across roles. Regression testing in hybrid settings emerges as a socio-technical system in which tools and practices co-evolve: automation and management platforms mediate collaboration, while human actors adapt workflows to bridge communication gaps and infrastructure constraints. The overall impact is a dual one—greater procedural discipline and autonomy coexisting with new coordination demands. Thus, the adoption of remote and hybrid work has transformed regression testing into a more traceable, technology-enabled, and learning-centered activity that sustains software quality through adaptive human–tool interaction.

\subsection{Member Checking}
\label{sec:memberchecking}

To strengthen the validity of the findings, a member checking procedure was conducted with seventeen practitioners who reviewed and rated the study’s results. Overall, the responses demonstrated strong convergence between the participants’ perspectives and the themes identified in the analysis. In relation to the regression testing process, nearly all respondents agreed with the three-phase structure of \textit{planning}, \textit{execution}, and \textit{closure}, confirming that the described sequence accurately reflects industry practice. Activities such as scope definition, entry criteria analysis, bug logging, and report preparation received the highest levels of agreement, while minor variations were attributed to organizational differences in workflow maturity. Tool-related findings were likewise validated: participants emphasized that compatibility, functionality, and integration with existing pipelines are decisive for regression testing effectiveness, although the perceived acceleration of testing depends on team experience and configuration stability. Regarding the remote and hybrid context, respondents reaffirmed that assertive communication and sustained focus are critical enablers of regression testing under distributed conditions, whereas availability delays and infrastructure limitations remain recurring challenges. Collectively, these results confirm that the interpretations drawn from the interviews are credible, contextually accurate, and representative of current regression testing practices in remote and hybrid environments.

\section{Discussion}
\label{sec:discussion}

This section interprets the findings in light of existing research. We first compare our results with prior work on regression testing and hybrid software development, then we highlight the implications for research and practice, and finally, we discuss threats to validity.

\subsection{Comparison with the Literature}

While prior research on remote and hybrid work in software engineering has primarily emphasized organizational or human factors such as communication, trust, and team cohesion \cite{ralph2020pandemic,jansen2024remote,smite2022future,de2022grounded}, few studies have investigated how these work arrangements reshape the technical activities of software development. The present study addresses this gap by relating distributed work directly to a technical and process-oriented practice: regression testing. By doing so, it provides a socio-technical perspective that links collaboration dynamics to concrete testing workflows, revealing how remote communication constraints, tool adoption, and automation strategies converge in maintaining software quality.

Our study expands prior research by connecting two areas that are often examined separately: regression testing and distributed software development. Previous studies describe the regression testing life cycle as a structured sequence of activities, including requirement analysis, planning, test design, execution, and defect tracking, aimed at verifying system stability after code modifications \cite{tomar2022regression,rahmani2021empirical,wang2023test,rosero2022software,greca2023state}. The findings of this work confirm that these fundamental stages remain present in remote and hybrid environments but show how they are reconfigured to support asynchronous coordination and communication across distributed teams. The process phases identified in this study, planning, execution, and closure, align with those described in traditional models, but they now rely more heavily on documentation, digital tools, and traceability to compensate for the lack of co-located interaction.

Our findings also reinforce existing discussions regarding tool selection and integration as persistent challenges in testing practice \cite{minhas2023checklists,wang2023test,rosero2022software}. Participants highlighted that the effectiveness of tools such as JIRA, Zephyr, and Azure DevOps depends on their compatibility with existing pipelines and the team’s technical maturity. While these tools improve visibility, traceability, and reporting, they also introduce configuration overhead, learning curves, and integration issues. This tension reflects patterns already observed in the literature, indicating that automation gains are often counterbalanced by coordination complexity.

With respect to remote collaboration, our results converge with findings that communication barriers are among the main difficulties faced by remote and hybrid software teams \cite{ralph2020pandemic,jackson2022collaboration,nguyen2024work,christensen2025hybrid,de2024post}. Participants described time-zone delays, asynchronous misunderstandings, and conflicting information among roles, but they also reported compensatory strategies such as increased documentation and process standardization. The present study, therefore, provides empirical support for the argument that remote and hybrid work introduces both friction in communication practices, making regression testing simultaneously more constrained and more disciplined.

The novelty of this study lies in its focus on regression testing within remote and hybrid software development environments, a topic that has received limited empirical attention. The results show that hybrid conditions shape the way regression testing is planned, executed, and managed, introducing both opportunities and challenges to maintain software reliability. Hybrid work supports continuous regression testing by enabling autonomy, uninterrupted execution, and comprehensive documentation, yet it also increases coordination effort, slows feedback cycles, and increases dependence on automation tools and digital infrastructure. Effective regression testing in hybrid settings requires mature processes, integrated toolchains, and structured communication to preserve traceability and consistency across distributed teams. When these elements are weak or inconsistently applied, hybrid arrangements can fragment collaboration and reduce testing efficiency. Overall, our findings suggest that the success of regression testing in hybrid environments depends on how well socio-technical practices are aligned to support quality assurance under distributed conditions.

\subsection{Implications for Research}

From a research perspective, this study advances understanding of how socio-technical factors shape quality assurance in distributed contexts. It provides an integrated model of regression testing that connects processes, tools, and human coordination mechanisms, showing that remote and hybrid work influence all three dimensions simultaneously. These results indicate that future studies should examine not only tool performance or fault detection rates but also the collaborative structures and communication strategies that sustain distributed testing. The methodological design used here, qualitative interviews complemented by member checking, offers a replicable approach for investigating other forms of testing, such as integration or acceptance testing, under hybrid arrangements. In addition, the evidence contributes to theory building on socio-technical adaptation, revealing how testing practices evolve through interaction between human agency and digital infrastructure.

\subsection{Implications for Practice}

For practitioners, the results emphasize the value of formalizing regression testing activities through structured planning and clear documentation. The three-phase model derived from this study can serve as a reference for teams operating remotely or adopting hybrid workflows. Managers and QA leads should invest in integrated test management platforms that unify planning, execution, and reporting, thereby ensuring transparency and consistency across distributed contributors. Establishing explicit communication protocols, such as standard reporting formats, scheduled updates, and defined escalation paths, can help mitigate coordination delays and information gaps. Furthermore, regression testing can be leveraged as an onboarding and learning mechanism for new team members, supporting knowledge transfer and familiarization with system behavior. Practitioners should recognize that while hybrid work enhances focus and autonomy, it also demands reliable infrastructure and deliberate attention to maintaining shared awareness among testers and developers.

\subsection{Threats to Validity}

As a qualitative investigation, the results are context-dependent and not intended for statistical generalization. The study’s purpose was to explore how regression testing adapts under remote and hybrid conditions, rather than to measure quantitative performance indicators. Internal validity was strengthened through participant diversity in terms of organization, experience, and geography, and by conducting a member checking phase with seventeen practitioners to confirm interpretive accuracy \cite{santos2017member}. Due to the nature of the ethics agreement governing this research, the full interview transcriptions cannot be publicly shared, as participants frequently referenced identifiable company names, colleagues, and project details. However, representative quotations are included throughout the text to illustrate the findings and provide transparency and credibility to the interpretations. Researcher interpretation may nonetheless introduce bias, and the absence of triangulation through complementary sources such as repository mining or observation constitutes a limitation. External validity is constrained to contexts comparable to those of the participating organizations, although the consistency between these findings and prior studies enhances their transferability \cite{ralph2020empirical,lenberg2024qualitative,de2024post}.

\section{Conclusions and Future Work}
\label{sec:conclusions}

This study investigated how regression testing is performed and experienced within remote and hybrid software development teams. The analysis showed that the fundamental phases of regression testing—planning, execution, and closure—remain consistent with traditional models but have been adapted to support distributed collaboration. Documentation, traceability, and tool integration have become essential mechanisms for sustaining quality assurance in the absence of co-located coordination. These adaptations confirm that regression testing continues to play a central role in maintaining system stability, but its execution now relies on digital infrastructure and deliberate communication practices that enable asynchronous teamwork. The findings indicate that remote and hybrid work amplify both the opportunities and the challenges of software testing. Automation and collaboration tools enhance visibility and efficiency, yet they also introduce additional coordination effort and require alignment across teams, time zones, and technologies. Regression testing has evolved into a socio-technical process shaped by the interaction between human coordination and technological mediation. This evolution reflects a broader transformation in software engineering, where quality assurance increasingly depends on documentation, shared accountability, and collective awareness within distributed teams. In our future work, we are going to extend this research by combining qualitative and quantitative approaches to study how regression testing effectiveness varies across organizations, levels of automation, and hybrid configurations. We are also going to conduct repository and pipeline analyses to complement the interview data and validate the adaptation patterns identified in this study. Furthermore, we are going to investigate whether the socio-technical changes observed here also occur in other testing practices, such as integration or acceptance testing. Through these next steps, we aim to deepen understanding of how hybrid work conditions reshape software quality processes and to provide actionable guidance for improving coordination, tool interoperability, and testing reliability in distributed environments.

\bibliographystyle{ACM-Reference-Format}
\bibliography{bibliography}

\appendix

\end{document}